\documentclass[aps,prb,preprint,superscriptaddress,floatfix,nobibnotes, showkeys]{revtex4-1}

\usepackage{graphicx}
\usepackage{amssymb}
\usepackage{amsmath}
\usepackage{bm} 

\usepackage{here}

\makeatletter 
\renewcommand{\thetable}{\arabic{table}} 
\renewcommand{\fnum@figure}{\textbf{Figure~\thefigure}}
\renewcommand{\fnum@table}{\textbf{Table~\thetable}}
\makeatother

\begin{document}


\title{Anisotropic spin-density distribution and magnetic anisotropy of 
strained La$_{1-x}$Sr$_{x}$MnO$_{3}$ thin films: 
Angle-dependent x-ray magnetic circular dichroism\\
\textmd{
(Running title: Anisotropic spin density in La$_{1-x}$Sr$_{x}$MnO$_{3}$ thin films)
}
}


\author{Goro Shibata} 
\email[Email: ]{shibata@wyvern.phys.s.u-tokyo.ac.jp}
\thanks{\\Telephone: +81-3-5841-7621\\}
\affiliation{Department of Physics, University of Tokyo, Bunkyo-ku, Tokyo 
113-0033, Japan}

\author{Miho Kitamura}
\affiliation{Photon Factory, Institute of Materials Structure Science, High 
Energy Accelerator Research Organization (KEK), Tsukuba, Ibaraki 305-0801, 
Japan}

\author{Makoto Minohara}
\affiliation{Photon Factory, Institute of Materials Structure Science, High 
Energy Accelerator Research Organization (KEK), Tsukuba, Ibaraki 305-0801, 
Japan}

\author{Kohei Yoshimatsu}
\affiliation{Department of Physics, University of Tokyo, Bunkyo-ku, Tokyo 
113-0033, Japan}
\affiliation{Photon Factory, Institute of Materials Structure Science, High 
Energy Accelerator Research Organization (KEK), Tsukuba, Ibaraki 305-0801, 
Japan}

\author{Toshiharu Kadono}
\affiliation{Department of Physics, University of Tokyo, Bunkyo-ku, Tokyo 
113-0033, Japan}

\author{Keisuke Ishigami}
\affiliation{Department of Physics, University of Tokyo, Bunkyo-ku, Tokyo 
113-0033, Japan}

\author{Takayuki Harano}
\affiliation{Department of Physics, University of Tokyo, Bunkyo-ku, Tokyo 
113-0033, Japan}

\author{Yukio Takahashi}
\affiliation{Department of Physics, University of Tokyo, Bunkyo-ku, Tokyo 
113-0033, Japan}

\author{Shoya Sakamoto}
\affiliation{Department of Physics, University of Tokyo, Bunkyo-ku, Tokyo 
113-0033, Japan}

\author{Yosuke Nonaka}
\affiliation{Department of Physics, University of Tokyo, Bunkyo-ku, Tokyo 
113-0033, Japan}

\author{Keisuke Ikeda}
\affiliation{Department of Physics, University of Tokyo, Bunkyo-ku, Tokyo 
113-0033, Japan}

\author{Zhendong Chi}
\affiliation{Department of Physics, University of Tokyo, Bunkyo-ku, Tokyo 
113-0033, Japan}

\author{Mitsuho Furuse}
\affiliation{National Institute of Advanced Industrial Science and Technology 
(AIST), Tsukuba, Ibaraki 305-8568, Japan}

\author{Shuichiro Fuchino}
\affiliation{National Institute of Advanced Industrial Science and Technology 
(AIST), Tsukuba, Ibaraki 305-8568, Japan}

\author{Makoto Okano}
\affiliation{National Institute of Advanced Industrial Science and Technology 
(AIST), Tsukuba, Ibaraki 305-8568, Japan}

\author{Jun-ichi Fujihira}
\affiliation{Fujihira Co., Ltd., Tsukuba, Ibaraki 305-0047, Japan}

\author{Akira Uchida}
\affiliation{Fujihira Co., Ltd., Tsukuba, Ibaraki 305-0047, Japan}

\author{Kazunori Watanabe}
\affiliation{Fujihira Co., Ltd., Tsukuba, Ibaraki 305-0047, Japan}

\author{Hideyuki Fujihira}
\affiliation{Fujihira Co., Ltd., Tsukuba, Ibaraki 305-0047, Japan}

\author{Seiichi Fujihira}
\affiliation{Fujihira Co., Ltd., Tsukuba, Ibaraki 305-0047, Japan}

\author{Arata Tanaka}
\affiliation{Department of Quantum Matter, Graduate School of Advanced Sciences of Matter, 
Hiroshima University, Higashi-Hiroshima, Hiroshima 739-8530, Japan}

\author{Hiroshi Kumigashira}
\affiliation{Photon Factory, Institute of Materials Structure Science, High 
Energy Accelerator Research Organization (KEK), Tsukuba, Ibaraki 305-0801, 
Japan}

\author{Tsuneharu Koide}
\affiliation{Photon Factory, Institute of Materials Structure Science, High 
Energy Accelerator Research Organization (KEK), Tsukuba, Ibaraki 305-0801, 
Japan}

\author{Atsushi Fujimori}
\affiliation{Department of Physics, University of Tokyo, Bunkyo-ku, Tokyo 
113-0033, Japan}


\date{\today}

\maketitle 
\newpage 

\begin{flushleft}

\section*{Abstract and Keywords}
\textbf{Magnetic anisotropies of ferromagnetic thin films are induced by 
epitaxial strain from the substrate via strain-induced anisotropy 
in the orbital magnetic moment and that in the spatial distribution of spin-polarized electrons. 
However, the preferential orbital occupation in ferromagnetic metallic 
La$_{1-x}$Sr$_{x}$MnO$_3$ (LSMO) thin films 
studied by x-ray linear dichroism (XLD) 
has always been found out-of-plane for both tensile and compressive epitaxial strain 
and hence irrespective of the magnetic anisotropy. 
In order to resolve this mystery, 
we directly probed the \textit{preferential orbital occupation of spin-polarized electrons} 
in LSMO thin films under strain 
by angle-dependent x-ray magnetic circular dichroism (XMCD). 
Anisotropy of the spin-density distribution was found to be in-plane 
for the tensile strain and out-of-plane for the compressive strain, 
consistent with the observed magnetic anisotropy. 
The ubiquitous out-of-plane preferential orbital occupation seen by XLD is 
attributed to the occupation of both spin-up and spin-down 
out-of-plane orbitals in the surface magnetic dead layer. 
}

Keywords: Electronic properties and materials; Ferromagnetism; Surfaces, interfaces and thin films; Magnetic properties and materials





\section*{Introduction}

Magnetic anisotropy is one of the most important properties of ferromagnets 
and its external control has been a major challenge both from 
the fundamental and applied science points of view \cite{Coey}. 
From the application point of view, 
enhancement of the magnetic anisotropy is necessary to realize 
magnets with high coercive fields, which can be utilized as 
high-density energy-storage magnets. 
From the scientific point of view, 
elucidating the microscopic origin of magnetic anisotropy has been an important issue 
because it is generally governed by the complex interplay between 
spin-orbit interaction and microscopic electronic states such as 
spin and orbital magnetic moments, band structures, 
and anisotropy of charge/spin densities. 
Especially, the magnetic anisotropy of ferromagnetic thin films is of great interest and importance 
because it can be controlled, e.g., by changing 
epitaxial strain and film thickness. 

As for oxide materials, the perovskite-type manganese oxide La$_{1-x}$Sr$_x$MnO$_3$ (LSMO) 
has been the most extensively studied ferromagnet due to its intriguing physical properties 
such as colossal magnetoresistance (CMR) and half-metallicity. 
The physical properties of LSMO can be controlled in various ways, e.g., by changing 
hole concentration $x$, temperature $T$, and external magnetic field $H$ (Ref.\ \onlinecite{IFT}). 	
In the case of thin films, their properties are also strongly affected by 
epitaxial strain which originates from the lattice mismatch between the film and the substrate. 
For example, Konishi \textit{et al.} \cite{Konishi} 
have shown that ferromagnetic (FM) metallic LSMO ($x$ = 0.3-0.5) thin films 
enter the A-type antiferromagnetic (AFM) metallic phase under tensile strain from a SrTiO$_3$ (STO) (001) substrate 
and the C-type AFM insulating phase under compressive strain from a LaAlO$_3$ (LAO) (001) substrate. 
The magnetic anisotropy of the LSMO thin films also depends on the epitaxial strain: 
the magnetic easy axes are in-plane when grown on the STO substrate and 
out-of-plane when grown on the LAO substrate
\cite{Tsui, KwonJMMM97}. 
First-principles calculations have predicted that 
the $d_{x^2-y^2}$ orbital is preferentially occupied under the tensile strain and that 
the $d_{3z^2-r^2}$ orbital is preferentially occupied under the compressive strain \cite{Konishi}. 
However, previous x-ray linear dichroism (XLD) experiments have shown that 
the $d_{3z^2-r^2}$ orbital is preferentially occupied for both 
STO and LAO substrates \cite{TebanoPRL08, ArutaPRB09, PesqueraNatCom12}. 
This apparent discrepancy with theory has been ascribed to the different orbital occupation 
between the surface and the bulk, that is, 
the spatial symmetry breaking at the surface leads to the preferential occupation 
of the $d_{3z^2-r^2}$ orbital \cite{TebanoPRL08, ArutaPRB09, PesqueraNatCom12}. 
Thus, the microscopic electronic and magnetic states of LSMO thin films and their relationship 
with the macroscopic magnetic properties have remained elusive so far. 

In the present work, we have employed a method which directly probes 
the orbital occupation of spin-polarized electrons 
using angle-dependent x-ray magnetic circular dichroism (XMCD) in 
core-level x-ray absorption spectroscopy (XAS). 
In the XMCD spin sum rule \cite{SpinSum}, 
in addition to the well-known term which represents 
the spin magnetic moment $\bm{M}_{\text{spin}}$, 
there is an additional term called `magnetic dipole term' $\bm{M}_{\text{T}}$ 
which represents the spatial anisotropy of spin-density distribution, namely, 
the orbital shapes of the spin-polarized electrons. 
While XLD is sensitive to the orbital polarization of \textit{all} the valence electrons, 
XMCD is sensitive to the orbital polarization of 
only \textit{spin-polarized} electrons and, therefore, 
one can directly probe the orbital states of electrons which contribute to the ferromagnetism. 
In general, it is difficult to deduce $\bm{M}_{\text{spin}}$ and $\bm{M}_{\text{T}}$ 
separately from a single XMCD spectrum by using the sum rule. 
However, as we shall see below, one can separate the magnetic moment into the 
$\bm{M}_{\text{spin}}$ and $\bm{M}_{\text{T}}$ components 
from the angular dependence of the XMCD spectra, 
because they have different angular dependencies 
\cite{TXMCD_Stohr, TXMCD_Durr, vanderLaanJPCM98, vanderLaanPRB98}. 
Hence, the spatial anisotropy of the spin-polarized electrons 
in the ferromagnetic materials can be deduced in addition to the total spin magnetic moment. 
Especially, in the geometry where $\bm{M}_{\text{spin}}$ is perpendicular 
to the incident x rays [so-called transverse XMCD (TXMCD) geometry] \cite{TXMCD_Durr}, 
one can extract the pure $\bm{M}_{\text{T}}$ component.

Although TXMCD has been theoretically studied since two decades ago 
\cite{TXMCD_Stohr, TXMCD_Durr, vanderLaanJPCM98, vanderLaanPRB98, TXMCD_Maruyama}, 
there have been only few experimental reports 
\cite{Durr_Science97, TXMCD_Mamiya, vanderLaanPRL10, TXMCD_Koide} 
because the direction of the magnetic field is fixed parallel 
or nearly parallel to the incident x rays in conventional XMCD measurement systems. 
Recently, we have developed an apparatus for angle-dependent XMCD experiments 
using a vector-type magnet where the direction of the magnetic field 
can be rotated using two pairs of superconducting magnets \cite{Vector_Furuse}. 
In this paper, we report on the angle-dependent XMCD and TXMCD experiments 
on ferromagnetic LSMO ($x = 0.3$) thin films grown on STO and LAO substrates, 
and investigate the effect of epitaxial strain on the orbital states of 
spin-polarized electrons. 
We have revealed that the LSMO thin film under tensile (compressive) strain has 
$d_{x^2-y^2}$-like ($d_{3z^2-r^2}$-like) spin-density distribution, 
which is different from the charge-density distribution deduced from the XLD measurements. 
The origin of the difference between the spin- and charge-density distributions 
is attributed to the preferential occupation of both the spin-up and spin-down 
$d_{3z^2 - r^2}$ orbitals at the surface, 
which suggests the formation of magnetic dead layers at the surface. 

\section*{Results}

\subsection*{Angular dependence of XMCD spectra and TXMCD.} 

Figure \ref{T_XMCD}a shows a schematic drawing of the experimental setup 
for angle-dependent XMCD. 
One can change the direction of the external magnetic field 
using two sets of superconducting magnets orthogonally arranged. 
The experimental geometry is schematically drawn in Fig.\ \ref{T_XMCD}b 
with the definition of the angles of incident x rays ($\theta_{\text{inc}}$), 
applied magnetic field ($\theta_{\bm{H}}$), 
and magnetization ($\theta_{\bm{M}}$). 
Note that in general $\theta_{\bm{M}}$ is not equal to $\theta_{\bm{H}}$ 
unless the applied magnetic field is large enough to fully 
align all the electron spins along the magnetic field direction. 
According to the XMCD sum rules \cite{SpinSum, OrbSum}, 
the `effective' spin magnetic moment 
$\hat{\bm{P}} \cdot [\bm{M}_{\text{spin}} + (7/2)\bm{M}_{\text{spin}}]$ 
is proportional to $\Delta I_3 + 2\Delta I_2$, 
where $\hat{\bm{P}}$ is a unit vector along the x-ray incident direction, 
and $\Delta I_3$ and $\Delta I_2$ are the integrals of the XMCD spectra 
over the Mn $L_3$ ($2p_{3/2} \to 3d$) and Mn $L_2$ ($2p_{1/2} \to 3d$) absorption edges, respectively. 
Under the assumption that 
the orbital magnetic moment $\bm{M}_{\text{orb}}$ and the magnetic dipole moment $\bm{M}_{\text{T}}$ 
are small enough compared to $\bm{M}_{\text{spin}}$, 
the projected spin magnetic moment $\hat{\bm{P}} \cdot \bm{M}_{\text{spin}}$ is 
approximately proportional to the XMCD integrals $\Delta I_3$ or $\Delta I_2$.
(For more information about angle-dependent XMCD, see Supplementary Fig.\ 1 and Supplementary Note 1.) 
In the present study, $\theta_{\text{inc}}$ was fixed to $45^{\circ}$ and 
$\theta_{\bm{M}}$ was varied through varying $\theta_{\bm{H}}$. 

We have grown LSMO ($x = 0.3$) thin films on the Nb-doped STO (tensile strain) and LAO (compressive strain) substrates 
by the laser molecular beam epitaxy method 
(See `Methods' section for the detail of sample preparation, 
and Supplementary Figs.\ 2-4 and Supplementary Note 2 for sample characterization.) 
Figures \ref{adXMCD}a and \ref{adXMCD}b show the Mn $L_{2,3}$-edge ($2p \to 3d$) XAS spectra 
of the LSMO thin films grown on the STO and LAO substrates, respectively, 
taken at $\theta_{\bm{H}}=45^{\circ}$ 
(where the magnetic field is applied parallel to the incident x rays). 
Since the spectral line shape of XAS was almost independent of $\theta_{\bm{H}}$ (see Supplementary Fig.\ 5), 
only the XAS spectra for $\theta_{\bm{H}}=45^{\circ}$ are shown here. 
The spectral line shape of XAS is similar to those obtained 
in previous XMCD studies of bulk \cite{LSMO_Koide} and 
thin-film \cite{ArutaPRB09, LSMO_Shibata} samples, 
and absorption signals of extrinsic Mn$^{2+}$ (Ref.\ \onlinecite{Mn2plus}) are hardly observed. 
Figures \ref{adXMCD}c and \ref{adXMCD}d show the Mn $L_{2,3}$-edge XMCD spectra of both the substrates 
for various $\theta_{\bm{H}}$'s. 
Systematic changes in the XMCD integrals at the Mn $L_3$ edge (approximately proportional to $\hat{\bm{P}} \cdot \bm{M}_{\text{spin}}$) can be seen which arise from 
the change in the magnetization direction $\theta_{\bm{M}}$ 
under varying $\theta_{\bm{H}}$. 
The XMCD integrals at the Mn $L_3$ edge reverse in sign around 
$\theta_{\bm{H}} = -15^{\circ}\; \mathchar`- -20^{\circ}$ for the LSMO/STO film and around 
$\theta_{\bm{H}} = -50^{\circ}\; \mathchar`- -55^{\circ}$  for the LSMO/LAO film. 
This means that the magnetization is directed nearly perpendicular to the incident x rays 
($\hat{\bm{P}} \cdot \bm{M}_{\text{spin}} \sim \hat{\bm{P}} \cdot \bm{M} \sim 0$) around these $\theta_{\bm{H}}$'s, namely, 
the TXMCD geometry is expected to exist around these angles. 

The orange and green curves in Fig.\ \ref{TXMCDkai}a shows the expanded XMCD spectra 
for LSMO/STO at $\theta_{\bm{H}}=-20^{\circ}$ and 
for LSMO/LAO at $\theta_{\bm{H}}=-50^{\circ}$, respectively. 
(We note that we have chosen these angles by the comparison with 
theoretical TXMCD spectra, as described below.)
Finite XMCD signals, which is expected to originate from 
the magnetic dipole term $\bm{M}_{\text{T}}$, are clearly observed. 
One may suspect that $\bm{M}_{\text{spin}}$ is not precisely aligned perpendicular to the x rays 
and yields this finite XMCD signals. 
This possibility, however, can be ruled out because the line shapes of the observed XMCD spectra 
are quite different from those of the conventional (longitudinal) XMCD (black curve in Fig. 3a). 
Furthermore, the spectral line shapes of LSMO/STO and LSMO/LAO are nearly identical 
but only the sign of the spectra is reversed. 
This suggests that the sign of $\bm{M}_{\text{T}}$, namely the anisotropy of the spin-density distribution, is reversed 
reflecting the opposite epitaxial strain. 

In order to show that the obtained spectra arise from genuine TXMCD, 
we have calculated the TXMCD spectra under tensile or compressive strain 
using the Mn$^{3+}$O$_6$ cluster model with $D_{4h}$ symmetry (see `Method' section for details). 
Here, only the Mn$^{3+}$ ($d^4$) valence state has been considered 
since the anisotropy of the charge/spin density is negligible 
for the Mn$^{4+}$ ($d^3$) valence state, 
where the $t_{2g \uparrow}$ levels are fully occupied and the $e_{g \uparrow}$ levels are empty. 
Using the parameter values listed in Supplementary Table 1, 
we have calculated the TXMCD spectra 
corresponding to both tensile and compressive strain, 
as shown in Fig.\ \ref{TXMCDkai}b. 
The calculated TXMCD spectra well reproduce the experimental ones, 
suggesting that the experimentally obtained spectra 
at $\theta_{\bm{H}}=-20^{\circ}$ for LSMO/STO and 
at $\theta_{\bm{H}}=-50^{\circ}$ for LSMO/LAO are the genuine TXMCD signals 
which reflect the anisotropic spin density on the Mn atom. 
Comparing the signs of the experimental TXMCD spectra with the calculated ones, 
it is clearly demonstrated that the spin-density distribution of the Mn 3$d$ electrons in the LSMO/STO (LSMO/LAO) thin film is 
more $d_{x^2-y^2}$-like ($d_{3z^2-r^2}$-like), 
consistent with the expectation for the tensile and compressive epitaxial strain from the substrates. 

\subsection*{Quantitative estimate of magnetic anisotropy energy and anisotropic spin-density distribution.}

We have seen in Figs.\ \ref{adXMCD}c and \ref{adXMCD}d that the sign change of $\hat{\bm{P}} \cdot \bm{M}_\text{spin}$ occurs 
around $\theta_{\bm{H}} \simeq -20^{\circ}$ for the LSMO/STO film and $\theta_{\bm{H}} \simeq -50^{\circ}$ for the LSMO/LAO film. 
If there were no magnetic anisotropy, 
$\theta_{\bm{M}}$ should be equal to $\theta_{\bm{H}}$ and the sign change 
should occur around $\theta_{\bm{H}} = -45^{\circ}$, where the incident x-ray beam is 
perpendicular to the magnetic field. 
The deviation of the sign change angle from 
$\theta_{\bm{H}} = -45^{\circ}$ in the present experiment indicates that 
$\theta_{\bm{M}}$ is not equal to $\theta_{\bm{H}}$ 
due to finite magnetic anisotropy. 
This offers the possibility to deduce the sign and magnitude of the magnetic anisotropy 
by fitting the measured angular dependence of the XMCD intensity 
to the theoretical one which incorporates the effect of magnetic anisotropy. 

Figures \ref{adXMCD2}a and \ref{adXMCD2}b 
show the $\theta_{\bm{H}}$ dependence of the projected effective spin magnetic moment 
$\hat{\bm{P}} \cdot \bm{M}_\text{spin}^\text{eff}\ (\equiv \hat{\bm{P}} \cdot [\bm{M}_\text{spin} + (7/2)\bm{M}_\text{T}] \simeq  \hat{\bm{P}} \cdot \bm{M}_\text{spin})$ 
obtained by applying the sum rule \cite{SpinSum} to the XMCD spectra in Figs.\ \ref{adXMCD}c (STO substrate) and \ref{adXMCD}d (LAO substrate), respectively. 
The obtained angular dependencies are different from the ones which assume $\theta_{\bm{H}} = \theta_{\bm{M}}$ (black dashed curves), 
indicating that the effect of magnetic anisotropy has to be taken into account. 
We have, therefore, simulated the obtained angular dependence of 
$\hat{\bm{P}} \cdot \bm{M}_\text{spin}^\text{eff}\ (\simeq \hat{\bm{P}} \cdot \bm{M}_\text{spin})$ 
based on the Stoner-Wohlfarth model \cite{SWmodel}. 
In this model, a single magnetic domain 
with uniaxial magnetic anisotropy of the lowest order 
(proportional to $\cos ^2 \theta_{\bm{M}}$) is assumed. 
Then, the magnetic energy (per volume) $E$ is given by an expression which contains $\theta_{\bm{M}}$. 
By minimizing $E$ with respect to $\theta_{\bm{M}}$ for each $\theta_{\bm{H}}$, 
one can deduce $\theta_{\bm{M}}$ as a function of $\theta_{\bm{H}}$, 
and can calculate the projected magnetic moment 
$\hat{\bm{P}} \cdot \bm{M}_\text{spin}^\text{eff} \equiv \hat{\bm{P}} \cdot [\bm{M}_\text{spin} + (7/2)\bm{M}_\text{T}]$ 
using the deduced $\theta_{\bm{M}}$. 
It is also possible to deduce the uniaxial magnetocrystalline anisotropy (MCA) constant $K_u$, the saturation magnetization $M_\text{sat}$, 
and the electric quadrupole moment $\langle Q_{zz} \rangle \equiv \langle 1-3\hat{z}^2 \rangle$ by taking these variables as parameters and 
fitting the simulated angular dependence to the experimental one 
(see the `Method' section for more details). 
The results of the simulations are shown in 
Figs.\ \ref{adXMCD2}a and \ref{adXMCD2}b 
by blue solid curves, 
showing good agreement with the experiment. 
The best-fit parameter values are listed in Table 1. 
The $K_u$ values in Table 1 ($K_u > 0$ corresponding to out-of-plane easy axis) 
clearly show that finite MCA is present in the LSMO/STO (LSMO/LAO) thin film 
which favors in-plane (out-of-plane) easy magnetization, 
consistent with the present (Supplementary Fig.\ 3) and previous \cite{Tsui, KwonJMMM97} magnetic measurements. 
Table 1 also shows that the electric quadrupole moment $\langle Q_{zz} \rangle = \langle 1 - 3\hat{z}^2 \rangle$ 
is positive (negative) for the STO (LAO) substrate. 
Since $(7/2)\langle Q_{zz} \rangle = +2$ for the $d_{x^{2}-y^{2}}$ orbital and $(7/2)\langle Q_{zz} \rangle = -2$ for the $d_{3z^{2}-r^{2}}$ orbital 
 (as shown in the first column of Fig.\ 1 in Ref.\ \onlinecite{TXMCD_Stohr}), 
the positive (negative) $\langle Q_{zz} \rangle$ for the LSMO/STO (LSMO/LAO) film implies that 
the charge distribution of spin-polarized electrons, 
namely, the distribution of the spin density, is more $x^2 - y^2$-like 
for the STO (tensile) substrate and more $3z^2 - r^2$-like for the LAO (compressive) 
substrate. 
This supports the TXMCD result that the spin-density distribution 
in the strained LSMO thin films is anisotropic. 
The degrees of the preferential orbital polarization 
$\left| (7/2)\langle Q_{zz} \rangle/2 \right|$ are estimated to be 
$\sim 2.5\%$ and $\sim 6\%$ for LSMO/STO and LSMO/LAO, respectively.

The advantage of the present method in deducing the magnetic anisotropy from the angle-dependent XMCD 
is that one can eliminate the effect of extrinsic spectral changes due to the saturation effect \cite{TEYdepth}, 
because the incident angle of the x rays is fixed. 
In addition, this method can be used in principle for 
dilute magnetic systems such as ultrathin films and lightly-doped magnetic semiconductors, 
for which the conventional magnetometry is hardly applicable, 
offering the possibility of estimating the magnetic anisotropy of these systems more accurately. 

\section*{Discussion}

The deduced anisotropic spin distribution in the LSMO thin films 
($x^{2}-y^{2}$-like in the case of the STO substrate and $3z^{2}-r^{2}$-like in the case of the LAO substrate) 
is consistent with the preferential orbital occupation expected from the strain from the substrate. 
It is also consistent with the preferential orbital occupation 
which has been suggested by the transport and magnetic measurements and 
the density-functional calculation 
\cite{Konishi}. 
On the other hand, the results of XLD measurements \cite{TebanoPRL08} 
show that the $d_{3z^2-r^2}$ orbital is more preferentially occupied than the $d_{x^2-y^2}$ orbital 
even in the case of tensile strain (STO substrate), 
which has been attributed to the symmetry breaking at the surface and interface \cite{PesqueraNatCom12}. 
The reason why the preferential orbital occupation seen by XMCD is consistent 
with that expected from the strain, in spite of its surface sensitivity comparable to XLD, 
may become apparent if one notices that 
XMCD is sensitive only to the \textit{spin-polarized} electrons 
while XLD is sensitive to \textit{all} the $d$ electrons. 
If the majority part of the surface Mn atoms occupies the $d_{3z^2-r^2}$ orbital due to the symmetry-breaking effect 
but are not spin-polarized, the $3z^2-r^2$-like charge-density distribution at the surface and interface should be observed in the XLD measurements, 
while the $x^2-y^2$-like spin-density distribution from underneath layers should be observed in the XMCD measurements. 
Indeed, there have been several reports which suggest the presence of 
magnetic dead layers at the surface or the interface of the 
FM LSMO thin films \cite{LSMO_Yoshimatsu, LSMO_Huijben}. 
The present angle-dependent XMCD and TXMCD studies, therefore, indicate close connection 
between the magnetic dead layer and the $3z^2-r^2$-like preferential orbital occupation at the surface of LSMO thin films. 
Further experiment is needed in order to test this hypothesis, e.g., by XLD measurements in the fluorescence-yield mode, 
in which we expect similar orbital polarization as the present study due to the longer penetration depth of the fluorescence-yield mode 
than the electron-yield mode.

\section*{Methods} 

\subsection*{Sample preparation.}
LSMO ($x=0.3$) thin films were grown on Nb-doped STO (001) and undoped LAO (001) (in the pseudo-cubic notation) substrates 
by laser molecular beam epitaxy \cite{LaserMBE}. 
Since the lattice constant of bulk LSMO is smaller (larger) than that of STO (LAO), 
the film is supposed to be under tensile (compressive) strain from the STO (LAO) substrate. 
The thickness of the thin films was around 100 unit cells ($\sim 40\ \text{nm}$) for both the samples. 
The growth rate was estimated from the 
intensity oscillation of the specular spot in 
reflection high-energy electron diffraction (RHEED) during the growth. 
The LSMO films were deposited at the temperature of $1050\ ^{\circ}\text{C}$ on the STO substrate and $650\ ^{\circ}\text{C}$ on the LAO substrate, 
under the oxygen pressure of $1 \times 10^{-4}$ Torr. 
Since the LSMO/LAO film tends to be fully relaxed at higher growth temperatures 
and be fully strained to become an antiferromagnetic insulator at lower growth temperatures \cite{Konishi}, 
we have adjusted the temperature so that the film is partially strained 
while the ferromagnetic metallicity of LSMO is maintained. 
After the growth of the films, both the samples were annealed 
at 400 $^{\circ}\text{C}$ for 45 minutes under 1 atm of O$_{2}$ 
to fill oxygen vacancies. 
The lattice constants of the films were evaluated by four-circle 
synchrotron x-ray diffraction (XRD) measurements 
at BL-7C of Photon Factory, 
High Energy Accelerator Research Organization (KEK-PF), 
and laboratory-based XRD measurements using the Cu $K \alpha$ line. 
The magnetization measurements were performed using a Quantum Design MPMS superconducting quantum interference device (SQUID) magnetometer. 
The temperature dependence of the resistivity was measured 
by the standard four-probe method. 
The results of the XRD, magnetization, and resistivity measurements 
are summarized in Supplementary Figs.\ 2-4 and Supplementary Note 2. 

\subsection*{XMCD measurements.}
The XAS and XMCD measurements were performed 
using a vector-magnet XMCD apparatus \cite{Vector_Furuse} 
with circularly polarized soft x rays 
at the helical undulator beam line BL-16A2 of KEK-PF. 
The measurement temperature $T$ for the LSMO/LAO film was 30 K, while it was set to 270 K for the LSMO/STO film. 
A lower $T$ was chosen for the LSMO/LAO film because the saturation magnetization at room temperature 
was low \cite{Konishi}, 
while a higher $T$ was chosen for the LSMO/STO film because the magnetic anisotropy at low temperature was too large 
to saturate the magnetization along the magnetic hard axis (out-of-plane direction).
The strength of the applied magnetic field was 0.7 T for the LSMO/STO film and 0.5 T for the LSMO/LAO film. 
The spectra were taken in the total electron-yield mode, 
which is a relatively surface-sensitive measurement mode 
(with a probing depth $\lambda$ of 
$\sim 3\ \text{nm}$) \cite{TEYdepth}. 
When the magnetic field is applied nearly parallel to the film surface, 
photo-ejected electrons are absorbed back to the sample 
due to the Lorentz force and the photocurrent drops to almost zero. 
In order to avoid this, we applied a negative bias voltage of $\sim 200\ \text{V}$ 
to the sample holder to help the photo-ejected electrons escape from the samples. 
The measurements were performed at a pressure of $\sim 1 \times 10^{-9}$ Torr. 
The intensity of the incident x rays was monitored by a photocurrent from the post-focusing mirror.

\subsection*{Cluster-model calculation.}
The cluster-model calculation was performed based on the method described in Ref.\ \onlinecite{TanakaCluster}, 
using the `Xtls' code (version 8.5) developed by Arata Tanaka. 
A distorted Mn$^{3+}$O$_6$ octahedral cluster with $D_{4h}$ symmetry (elongated or shrunk along the [001] direction) was used (Fig.\ \ref{TXMCDkai}c). 
The energy levels of the Mn 3$d$ orbitals under this symmetry are schematically drawn in Fig.\ \ref{TXMCDkai}d. 
The Mn 3$d$, Mn 2$p$ core, and O 2$p$ orbitals 
were taken as basis functions. 
Charge transfer from the ligand O $2p$ to the Mn $3d$ orbitals was taken into account, and 
we considered three electron configurations for both the initial and final states: 
$2p^6 3d^4$, 
$2p^6 3d^5 \underline{L}$, 
and 
$2p^6 3d^6 \underline{L}^2$ 
for the initial state, and 
$2p^5 3d^5$, 
$2p^5 3d^6 \underline{L}$, 
and 
$2p^5 3d^7 \underline{L}^2$ 
for the final state. 
We adjusted the following parameters to reproduce the experimental TXMCD spectra: 
$U_{dd}$ (Mn $3d$-$3d$ Coulomb energy), 
$U_{pd}$ (Mn $2p$-$3d$ Coulomb energy), 
$\Delta$ (charge-transfer energy from O $2p$ to Mn $3d$), 
$(pd \sigma)$ (Slater-Koster parameter between Mn $3d$ and O $2p$), 
and 
$10Dq$ (crystal-field splitting between the Mn $e_g$ and $t_{2g}$ levels). 
The magnitude of the $D_{4h}$ crystal-field splitting $8Cp$ (splitting between the $x^2-y^2$ and $3z^2-r^2$ levels) \cite{vanderLaanPRL10} 
was fixed to 0.08 eV and only its sign was varied, 
because varying the magnitude of $8Cp$ only changed the magnitude of XMCD and did not change the spectral line shape. 
We neglected the anisotropy of transfer integrals due to the $D_{4h}$ symmetry of the MnO$_6$ cluster 
and transfer integrals between the O $2p$ orbitals, 
in order to reduce the number of adjustable parameters. 
The x-ray incident angle was chosen to be in the [101] direction. 
In order to fully align the spins perpendicular to the incident x rays, 
a molecular field (an effective magnetic field corresponding to the exchange interaction) 
of 0.01 eV along the [$\bar{1}01$] direction was introduced. 
We note that 
this molecular field is strong enough to saturate the magnetization of the Mn ions. 

\subsection*{Simulation of angular dependence of $\hat{\bm{P}} \cdot \bm{M}_\text{spin}$ based on the Stoner-Wohlfarth model.} 

We have adopted the Stoner-Wohlfarth model \cite{SWmodel} in order to simulate the angular dependence of the projected effective spin magnetic moment 
$\hat{\bm{P}} \cdot \bm{M}_\text{spin}^\text{eff}\ (\simeq \hat{\bm{P}} \cdot \bm{M}_\text{spin})$ 
in Figs.\ \ref{adXMCD2}a and \ref{adXMCD2}b. 
By assuming that the film has only a single magnetic domain and that 
the magnetic anisotropy has only the uniaxial component of the lowest order, 
the magnetic energy (per volume) $E$ can be expressed as 
\begin{eqnarray}
E = & - &\mu_{0}M_{\text{sat}}H \cos(\theta_{\bm{M}} - \theta_{\bm{H}}) \nonumber \\
& + & \frac{\mu_{0}}{2}M_{\text{sat}}^{2} \cos^{2}\theta_{\bm{M}} - K_{\text{u}} \cos^{2} \theta_{\bm{M}}, 
\label{Eng}
\end{eqnarray}
where $H$ is the magnitude of the external magnetic field, $M_{\text{sat}}$ is the saturation magnetization, 
and $K_{\text{u}}$ is the uniaxial anisotropy constant for MCA ($K_u > 0$ for out-of-plane easy axis). 
The three terms in Eq.\ (\ref{Eng}) represent the Zeeman energy due to the 
applied magnetic field, the shape magnetic anisotropy which originates from the demagnetization field in the film, 
and the MCA which originates from a conbined effect of 
microscopic electron occupation and spin-orbit interaction. 
By minimizing $E$ with respect to $\theta_{\bm{M}}$, 
we deduced $\theta_{\bm{M}}$ as a function of $\theta_{\bm{H}}$, $H$, $K_{\text{u}}$, and $M_{\text{sat}}$. 
Then, the projection of the effective spin magnetic moment 
$\hat{\bm{P}} \cdot \bm{M}_\text{spin}^\text{eff} \equiv \hat{\bm{P}} \cdot [\bm{M}_\text{spin} + (7/2)\bm{M}_\text{T}]$ 
was calculated using the deduced $\theta_{\bm{M}}$ by the following equation: 
\begin{eqnarray}
& & \hat{\bm{P}} \cdot \bm{M}_{\text{spin}} + (7/2) \hat{\bm{P}} \cdot \bm{M}_{\text{T}} \nonumber \\
& = & M_{\text{sat}} \cos (\theta _{\bm{M}} - \theta _{\text{inc}}) + (7/4) \langle Q_{zz} \rangle M_{\text{sat}} (2 \cos \theta _{\bm{M}} \cos \theta _{\text{inc}} - \sin \theta _{\bm{M}} \sin \theta _{\text{inc}}), 
\label{angledep}
\end{eqnarray}
where $\langle Q_{zz} \rangle \equiv \langle 1 - 3\hat{z}^2 \rangle$ is the electric quadrupole moment 
[For the derivation of Eq.\ (\ref{angledep}), see Supplementary Note 1]. 
This gives the $\theta_{\bm{H}}$ dependence of the projected moment $\hat{\bm{P}} \cdot \bm{M}_\text{spin}^\text{eff}$ 
for a set of parameters ($K_u$, $M_{\text{sat}}$, and $\langle Q_{zz} \rangle$). 
The obtained  $\theta_{\bm{H}}$ dependence was fitted to the experimental one (Fig.\ \ref{adXMCD2}) 
to deduce $K_u$, $M_{\text{sat}}$, and $\langle Q_{zz} \rangle$ 
using the least-square method. 

\subsection*{Data availability}
The data supporting the findings of this study are available from the corresponding authors on reasonable request. 

\begin{acknowledgments} 
We would like to thank Kenta Amemiya, Masako Sakamaki, and Reiji Kumai 
for valuable technical support at KEK-PF. 
We would also like to thank Hiroki Wadati for providing us with information about 
the XLD studies of LSMO thin films. 
This work was supported by a Grant-in-Aid for Scientific Research from 
the JSPS (22224005, 15H02109, 15K17696, and 16H02115). 
The experiment was done under the approval of the Photon Factory 
Program Advisory Committee 
(proposal No.\ 2016S2-005, No.\ 2013S2-004, No.\ 2016G066, No.\ 2014G177, No.\ 2012G667, and 2015S2-005). 
Part of this work was performed using a SQUID magnetometer 
at the Cryogenic Research Center, the University of Tokyo. 
G.S.\ acknowledges support from 
Advanced Leading Graduate Course for Photon Science (ALPS) 
at the University of Tokyo and the JSPS Research Fellowships 
for Young Scientists (Project No. 26.11615). 
A.F.\ is an adjunct member of Center for Spintronics Research Network (CSRN), 
the University of Tokyo, under Spintronics Research Network of Japan (Spin-RNJ). 
\end{acknowledgments}

\section*{Competing financial interests:}
The authors declare no competing financial interests. 

\section*{Author contributions}
G.S., K.Y., T.Kadono, K.Ishigami, T.H., Y.T, S.S., Y.N., K.Ikeda, and Z.C. 
performed XMCD measurements with the assistance of T.Koide and A.F. 
M.K., M.M., and K.Y. grew and characterized thin films with the assistance of H.K. 
M.F., M.O., S.Fuchino, A.U., and J.-i.F. developed the vector-type superconducting magnet. 
J.-i.F., A.U., K.W., H.F., K.Ishigami, T.H., Y.N., 
T.Kadono, Y.T., S.S., K.Ikeda, Z.C., and G.S. 
were involved in the design, construction, and improvement of the XMCD measurement chamber, 
with assistance of S.Fujihira, T.Koide, and A.F. 
G.S. analyzed the XMCD data and performed the cluster-model calculation. 
A.T. developed the code for the cluster-model calculation (Xtls version 8.5). 
G.S. and A.F. wrote the manuscript with suggestions by M.K., M.M., K.Y., T.Koide, 
and all the other coauthors. 
A.F. was responsible for overall project direction and planning.



\begin{thebibliography}{10}

\section*{References}

\expandafter\ifx\csname url\endcsname\relax
  \def\url#1{\texttt{#1}}\fi
\expandafter\ifx\csname urlprefix\endcsname\relax\def\urlprefix{URL }\fi
\providecommand{\bibinfo}[2]{#2}
\providecommand{\eprint}[2][]{\url{#2}}

\bibitem{Coey}
\bibinfo{author}{Coey, J. M.~D.}
\newblock \emph{\bibinfo{title}{Magnetism and Magnetic Materials}}
  (\bibinfo{publisher}{Cambridge University Press}, \bibinfo{address}{New
  York}, \bibinfo{year}{2009}).

\bibitem{IFT}
\bibinfo{author}{Imada, M.}, \bibinfo{author}{Fujimori, A.} \&
  \bibinfo{author}{Tokura, Y.}
\newblock \bibinfo{title}{Metal-insulator transitions}.
\newblock \emph{\bibinfo{journal}{Rev. Mod. Phys.}}
  \textbf{\bibinfo{volume}{70}}, \bibinfo{pages}{1039--1263}
  (\bibinfo{year}{1998}).

\bibitem{Konishi}
\bibinfo{author}{Konishi, Y.} \emph{et~al.}
\newblock \bibinfo{title}{Orbital-state-mediated phase-control of manganites}.
\newblock \emph{\bibinfo{journal}{J. Phys. Soc. Jpn.}}
  \textbf{\bibinfo{volume}{68}}, \bibinfo{pages}{3790--3793}
  (\bibinfo{year}{1999}).

\bibitem{Tsui}
\bibinfo{author}{Tsui, F.}, \bibinfo{author}{Smoak, M.~C.},
  \bibinfo{author}{Nath, T.~K.} \& \bibinfo{author}{Eom, C.~B.}
\newblock \bibinfo{title}{Strain-dependent magnetic phase diagram of epitaxial
  La$_{0.67}$Sr$_{0.33}$MnO$_3$ thin films}.
\newblock \emph{\bibinfo{journal}{Appl. Phys. Lett.}}
  \textbf{\bibinfo{volume}{76}}, \bibinfo{pages}{2421--2423}
  (\bibinfo{year}{2000}).

\bibitem{KwonJMMM97}
\bibinfo{author}{Kwon, C.} \emph{et~al.}
\newblock \bibinfo{title}{Stress-induced effects in epitaxial (La$_{0.7}$Sr$_{0.3}$)MnO$_3$
  films}.
\newblock \emph{\bibinfo{journal}{J. Magn. Magn. Mater.}}
  \textbf{\bibinfo{volume}{172}}, \bibinfo{pages}{229 -- 236}
  (\bibinfo{year}{1997}).

\bibitem{TebanoPRL08}
\bibinfo{author}{Tebano, A.} \emph{et~al.}
\newblock \bibinfo{title}{Evidence of orbital reconstruction at interfaces in
  ultrathin La$_{0.67}$Sr$_{0.33}$MnO$_3$ 
  films}.
\newblock \emph{\bibinfo{journal}{Phys. Rev. Lett.}}
  \textbf{\bibinfo{volume}{100}}, \bibinfo{pages}{137401}
  (\bibinfo{year}{2008}).

\bibitem{ArutaPRB09}
\bibinfo{author}{Aruta, C.} \emph{et~al.}
\newblock \bibinfo{title}{Orbital occupation, atomic moments, and magnetic
  ordering at interfaces of manganite thin films}.
\newblock \emph{\bibinfo{journal}{Phys. Rev. B}} \textbf{\bibinfo{volume}{80}},
  \bibinfo{pages}{014431} (\bibinfo{year}{2009}).

\bibitem{PesqueraNatCom12}
\bibinfo{author}{Pesquera, D.} \emph{et~al.}
\newblock \bibinfo{title}{Surface symmetry-breaking and strain effects on
  orbital occupancy in transition metal perovskite epitaxial films}.
\newblock \emph{\bibinfo{journal}{Nat. Commun.}} \textbf{\bibinfo{volume}{3}},
  \bibinfo{pages}{1189} (\bibinfo{year}{2012}).

\bibitem{SpinSum}
\bibinfo{author}{Carra, P.}, \bibinfo{author}{Thole, B.~T.},
  \bibinfo{author}{Altarelli, M.} \& \bibinfo{author}{Wang, X.}
\newblock \bibinfo{title}{X-ray circular dichroism and local magnetic fields}.
\newblock \emph{\bibinfo{journal}{Phys. Rev. Lett.}}
  \textbf{\bibinfo{volume}{70}}, \bibinfo{pages}{694--697}
  (\bibinfo{year}{1993}).

\bibitem{TXMCD_Stohr}
\bibinfo{author}{St\"ohr, J.} \& \bibinfo{author}{K\"onig, H.}
\newblock \bibinfo{title}{Determination of spin- and orbital-moment
  anisotropies in transition metals by angle-dependent x-ray magnetic circular
  dichroism}.
\newblock \emph{\bibinfo{journal}{Phys. Rev. Lett.}}
  \textbf{\bibinfo{volume}{75}}, \bibinfo{pages}{3748--3751}
  (\bibinfo{year}{1995}).

\bibitem{TXMCD_Durr}
\bibinfo{author}{D\"urr, H.~A.} \& \bibinfo{author}{van~der Laan, G.}
\newblock \bibinfo{title}{Magnetic circular x-ray dichroism in transverse
  geometry: Importance of noncollinear ground state moments}.
\newblock \emph{\bibinfo{journal}{Phys. Rev. B}} \textbf{\bibinfo{volume}{54}},
  \bibinfo{pages}{R760--R763} (\bibinfo{year}{1996}).

\bibitem{vanderLaanJPCM98}
\bibinfo{author}{van~der Laan, G.}
\newblock \bibinfo{title}{Microscopic origin of magnetocrystalline anisotropy
  in transition metal thin films}.
\newblock \emph{\bibinfo{journal}{J. Phys. Condens. Matter}}
  \textbf{\bibinfo{volume}{10}}, \bibinfo{pages}{3239} (\bibinfo{year}{1998}).

\bibitem{vanderLaanPRB98}
\bibinfo{author}{van~der Laan, G.}
\newblock \bibinfo{title}{Relation between the angular dependence of magnetic
  x-ray dichroism and anisotropic ground-state moments}.
\newblock \emph{\bibinfo{journal}{Phys. Rev. B}} \textbf{\bibinfo{volume}{57}},
  \bibinfo{pages}{5250--5258} (\bibinfo{year}{1998}).

\bibitem{TXMCD_Maruyama}
\bibinfo{author}{Maruyama, T.}, \bibinfo{author}{Hojo, I.},
  \bibinfo{author}{Nagamatsu, S.-i.} \& \bibinfo{author}{Fujikawa, T.}
\newblock \bibinfo{title}{Theoretical study of angular-dependent $L_{2,3}$-edge
  XMCD}.
\newblock \emph{\bibinfo{journal}{J. Electron. Spectrosc. Relat. Phenom.}}
  \textbf{\bibinfo{volume}{180}}, \bibinfo{pages}{46--52}
  (\bibinfo{year}{2010}).

\bibitem{Durr_Science97}
\bibinfo{author}{D{\"u}rr, H.~A.} \emph{et~al.}
\newblock \bibinfo{title}{Element-specific magnetic anisotropy determined by
  transverse magnetic circular x-ray dichroism}.
\newblock \emph{\bibinfo{journal}{Science}} \textbf{\bibinfo{volume}{277}},
  \bibinfo{pages}{213--215} (\bibinfo{year}{1997}).

\bibitem{TXMCD_Mamiya}
\bibinfo{author}{{Mamiya}, K.} \emph{et~al.}
\newblock \bibinfo{title}{{Angle-resolved soft X-ray magnetic circular
  dichroism in a monatomic Fe layer facing an MgO(0 0 1) tunnel barrier}}.
\newblock \emph{\bibinfo{journal}{Radiat. Phys. Chem.}}
  \textbf{\bibinfo{volume}{75}}, \bibinfo{pages}{1872--1877}
  (\bibinfo{year}{2006}).

\bibitem{vanderLaanPRL10}
\bibinfo{author}{van~der Laan, G.}, \bibinfo{author}{Chopdekar, R.~V.},
  \bibinfo{author}{Suzuki, Y.} \& \bibinfo{author}{Arenholz, E.}
\newblock \bibinfo{title}{Strain-induced changes in the electronic structure of
  ${\mathrm{MnCr}}_{2}{\mathrm{O}}_{\mathrm{4}}$ thin films probed by x-ray
  magnetic circular dichroism}.
\newblock \emph{\bibinfo{journal}{Phys. Rev. Lett.}}
  \textbf{\bibinfo{volume}{105}}, \bibinfo{pages}{067405}
  (\bibinfo{year}{2010}).

\bibitem{TXMCD_Koide}
\bibinfo{author}{Koide, T.} \emph{et~al.}
\newblock \bibinfo{title}{Gigantic transverse x-ray magnetic circular dichroism
  in ultrathin Co in Au/Co/Au(001)}.
\newblock \emph{\bibinfo{journal}{J. Phys. Conf. Ser.}}
  \textbf{\bibinfo{volume}{502}}, \bibinfo{pages}{012002}
  (\bibinfo{year}{2014}).

\bibitem{Vector_Furuse}
\bibinfo{author}{Furuse, M.} \emph{et~al.}
\newblock \bibinfo{title}{HTS vector magnet for magnetic circular dichroism
  measurement}.
\newblock \emph{\bibinfo{journal}{IEEE Trans. Appl. Supercond.}}
  \textbf{\bibinfo{volume}{23}}, \bibinfo{pages}{4100704}
  (\bibinfo{year}{2013}).

\bibitem{OrbSum}
\bibinfo{author}{Thole, B.~T.}, \bibinfo{author}{Carra, P.},
  \bibinfo{author}{Sette, F.} \& \bibinfo{author}{van~der Laan, G.}
\newblock \bibinfo{title}{X-ray circular dichroism as a probe of orbital
  magnetization}.
\newblock \emph{\bibinfo{journal}{Phys. Rev. Lett.}}
  \textbf{\bibinfo{volume}{68}}, \bibinfo{pages}{1943--1946}
  (\bibinfo{year}{1992}).

\bibitem{LSMO_Koide}
\bibinfo{author}{Koide, T.} \emph{et~al.}
\newblock \bibinfo{title}{Close correlation between the magnetic moments,
  lattice distortions, and hybridization in ${\mathrm{LaMnO}}_{3}$ and
  ${\mathrm{La}}_{1-\mathit{x}}{\mathrm{Sr}}_{\mathit{x}}{\mathrm{MnO}}_{3+\delta{}}$: Doping-dependent magnetic circular x-ray dichroism study}.
\newblock \emph{\bibinfo{journal}{Phys. Rev. Lett.}}
  \textbf{\bibinfo{volume}{87}}, \bibinfo{pages}{246404}
  (\bibinfo{year}{2001}).

\bibitem{LSMO_Shibata}
\bibinfo{author}{Shibata, G.} \emph{et~al.}
\newblock \bibinfo{title}{Thickness-dependent ferromagnetic metal to
  paramagnetic insulator transition in
  ${\mathrm{La}}_{0.6}$${\mathrm{Sr}}_{0.4}$${\mathrm{MnO}}_{3}$ thin films
  studied by x-ray magnetic circular dichroism}.
\newblock \emph{\bibinfo{journal}{Phys. Rev. B}} \textbf{\bibinfo{volume}{89}},
  \bibinfo{pages}{235123} (\bibinfo{year}{2014}).

\bibitem{Mn2plus}
\bibinfo{author}{de~Jong, M.~P.} \emph{et~al.}
\newblock \bibinfo{title}{Evidence for ${\mathrm{Mn}}^{2+}$ ions at surfaces of
  ${\mathrm{La}}_{0.7}{\mathrm{Sr}}_{0.3}\mathrm{Mn}{\mathrm{O}}_{3}$ thin
  films}.
\newblock \emph{\bibinfo{journal}{Phys. Rev. B}} \textbf{\bibinfo{volume}{71}},
  \bibinfo{pages}{014434} (\bibinfo{year}{2005}).

\bibitem{SWmodel}
\bibinfo{author}{Stoner, E.~C.} \& \bibinfo{author}{Wohlfarth, E.~P.}
\newblock \bibinfo{title}{A mechanism of magnetic hysteresis in heterogeneous
  alloys}.
\newblock \emph{\bibinfo{journal}{Philos. Trans. R. Soc. London, Ser. A}}
  \textbf{\bibinfo{volume}{240}}, \bibinfo{pages}{599--642}
  (\bibinfo{year}{1948}).

\bibitem{TEYdepth}
\bibinfo{author}{Nakajima, R.}, \bibinfo{author}{St\"ohr, J.} \&
  \bibinfo{author}{Idzerda, Y.~U.}
\newblock \bibinfo{title}{Electron-yield saturation effects in $L$-edge x-ray
  magnetic circular dichroism spectra of Fe, Co, and Ni}.
\newblock \emph{\bibinfo{journal}{Phys. Rev. B}} \textbf{\bibinfo{volume}{59}},
  \bibinfo{pages}{6421--6429} (\bibinfo{year}{1999}).

\bibitem{LSMO_Yoshimatsu}
\bibinfo{author}{Yoshimatsu, K.}, \bibinfo{author}{Horiba, K.},
  \bibinfo{author}{Kumigashira, H.}, \bibinfo{author}{Ikenaga, E.} \&
  \bibinfo{author}{Oshima, M.}
\newblock \bibinfo{title}{Thickness dependent electronic structure of
  La$_{0.6}$Sr$_{0.4}$MnO$_3$ layer in SrTiO$_{3}$/La$_{0.6}$Sr$_{0.4}$MnO$_3$/SrTiO$_{3}$ heterostructures studied
  by hard x-ray photoemission spectroscopy}.
\newblock \emph{\bibinfo{journal}{Appl. Phys. Lett.}}
  \textbf{\bibinfo{volume}{94}}, \bibinfo{pages}{071901--1--3}
  (\bibinfo{year}{2009}).

\bibitem{LSMO_Huijben}
\bibinfo{author}{Huijben, M.} \emph{et~al.}
\newblock \bibinfo{title}{Critical thickness and orbital ordering in ultrathin
  ${\text{La}}_{0.7}{\text{Sr}}_{0.3}{\text{MnO}}_{3}$ films}.
\newblock \emph{\bibinfo{journal}{Phys. Rev. B}} \textbf{\bibinfo{volume}{78}},
  \bibinfo{pages}{094413} (\bibinfo{year}{2008}).

\bibitem{LaserMBE}
\bibinfo{author}{Horiba, K.} \emph{et~al.}
\newblock \bibinfo{title}{A high-resolution synchrotron-radiation
  angle-resolved photoemission spectrometer with \textit{in situ} oxide thin film growth
  capability}.
\newblock \emph{\bibinfo{journal}{Rev. Sci. Instrum.}}
  \textbf{\bibinfo{volume}{74}}, \bibinfo{pages}{3406--3412}
  (\bibinfo{year}{2003}).

\bibitem{TanakaCluster}
\bibinfo{author}{Tanaka, A.} \& \bibinfo{author}{Jo, T.}
\newblock \bibinfo{title}{Resonant 3$d$, 3$p$ and 3$s$ photoemission in transition
  metal oxides predicted at 2$p$ threshold}.
\newblock \emph{\bibinfo{journal}{J. Phys. Soc. Jpn.}}
  \textbf{\bibinfo{volume}{63}}, \bibinfo{pages}{2788--2807}
  (\bibinfo{year}{1994}).

\bibitem{XCrySDen}
\bibinfo{author}{Kokalj, A.}
\newblock \bibinfo{title}{XCrySDen---a new program for displaying crystalline
  structures and electron densities}.
\newblock \emph{\bibinfo{journal}{J. Mol. Graphics Modell.}}
  \textbf{\bibinfo{volume}{17}}, \bibinfo{pages}{176--179}
  (\bibinfo{year}{1999}).

\end{thebibliography}

\clearpage

\section*{Figure Legends:}

\begin{figure}[H]
\caption{
\textbf{Experimental geometry of angle-dependent 
x-ray magnetic circular dichroism (XMCD).} 
(\textbf{a}) Schematic drawing of the experimental setup. 
(\textbf{b}) Definition of the angles of incident x rays ($\theta _{\text{inc}}$), 
magnetic field ($\theta _{\bm{H}}$), 
and magnetization ($\theta _{\bm{M}}$).
$\theta _{\text{inc}}$ was fixed at $45^{\circ}$ in the present work. 
$\hat{\bm{P}}$ is a unit vector along the x-ray incident direction, 
which is defined to be antiparallel to the wavevector of x rays $\bm{k}$. 
}
\label{T_XMCD}
\end{figure}


\begin{figure}[H]
\caption{
\textbf{X-ray absorption spectroscopy (XAS) and angle-dependent XMCD spectra of the 
La$_{1-x}$Sr$_{x}$MnO$_3$ (LSMO, $x=0.3$) thin films at the Mn $L_{2,3}$ absorption edges.} 
(\textbf{a, b}) XAS spectra of the LSMO thin films grown on the Nb-doped SrTiO$_3$ (STO) (\textbf{a}) and 
LaAlO$_3$ (LAO) (\textbf{b}) substrates. 
The light red and light blue curves are the absorption spectra 
for the positive ($\sigma +$) and negative ($\sigma -$) helicity photons, respectively, 
and the green curves are the absorption spectra averaged over both the helicities. 
The spectra have been normalized so that 
the height of the averaged XAS spectra is equal to unity. 
(\textbf{c, d}) XMCD spectra of the LSMO/STO (\textbf{c}) and LSMO/LAO (\textbf{d}) thin films 
with varying $\theta_{\bm{H}}$. 
See Fig.\ \ref{T_XMCD} for the experimental geometry. 
}
\label{adXMCD}
\end{figure}


\begin{figure}[H]
\caption{
\textbf{Transverse XMCD (TXMCD).} 
(\textbf{a}) 
Experimental TXMCD spectra of the LSMO thin films 
on the STO (orange) and LAO (green) substrates 
compared with the longitudinal XMCD (LXMCD) spectra (black). 
Inset shows the schematic drawing of the TXMCD geometry. 
(\textbf{b}) Calculated TXMCD spectra based on the 
Mn$^{3+}$O$_6$ cluster model with $D_{4h}$ symmetry. 
(\textbf{c}) Schematic drawing of the Mn$^{3+}$O$_6$ cluster, 
(\textbf{d}) Schematic drawing of the energy levels of the Mn 3$d$ orbitals under $D_{4h}$ symmetry. 
Here, $Cp$ is a parameter proportional to the crystal-field splitting between the $x^2-y^2$ and $3z^2-r^2$ levels \cite{vanderLaanPRL10}, and 
$Cp=+0.01$ eV ($Cp=-0.01$ eV) corresponds to the case where the $x^2-y^2$ ($3z^2-r^2$) level has lower energy than the $3z^2-r^2$ ($x^2-y^2$) level. 
Panels \textbf{c} and \textbf{d} describe the case of $Cp < 0$. 
The parameter values used for the cluster-model calculation are listed in Supplementary Table 1. 
Panel \textbf{c} was drawn using XCrySDen \cite{XCrySDen}. 
}
\label{TXMCDkai}
\end{figure}


\begin{figure}[H]
\caption{
\textbf{Angular dependence of the projected magnetic moment.} 
(\textbf{a, b}) $\theta_{\bm{H}}$-dependencies of the projected effective spin magnetic moment 
$\hat{\bm{P}} \cdot \bm{M}_\text{spin}^\text{eff}\ (\sim \hat{\bm{P}} \cdot \bm{M}_\text{spin}$) 
deduced from the experimental data using the spin XMCD sum rule \cite{SpinSum} (circle) and 
its simulations. 
\textbf{a} and \textbf{b} are the data for the LSMO/STO and LSMO/LAO thin films, respectively. 
The black dashed curve describes the case where there is no magnetic anisotropy, and 
the blue solid curve describes the case where the shape magnetic anisotropy and magnetocrystalline anisotropy (MCA) are 
taken into account. 
Insets show the $\theta_{\bm{M}}$ vs $\theta_{\bm{H}}$ relations 
deduced from the simulation. 
The strength of the applied magnetic field was 0.7 T for the LSMO/STO film and 0.5 T for the LSMO/LAO film. 
See Fig.\ \ref{T_XMCD} for the experimental geometry. 
}
\label{adXMCD2}
\end{figure}


\clearpage
\setcounter{figure}{0}
\begin{figure*}
\includegraphics[width=16cm]{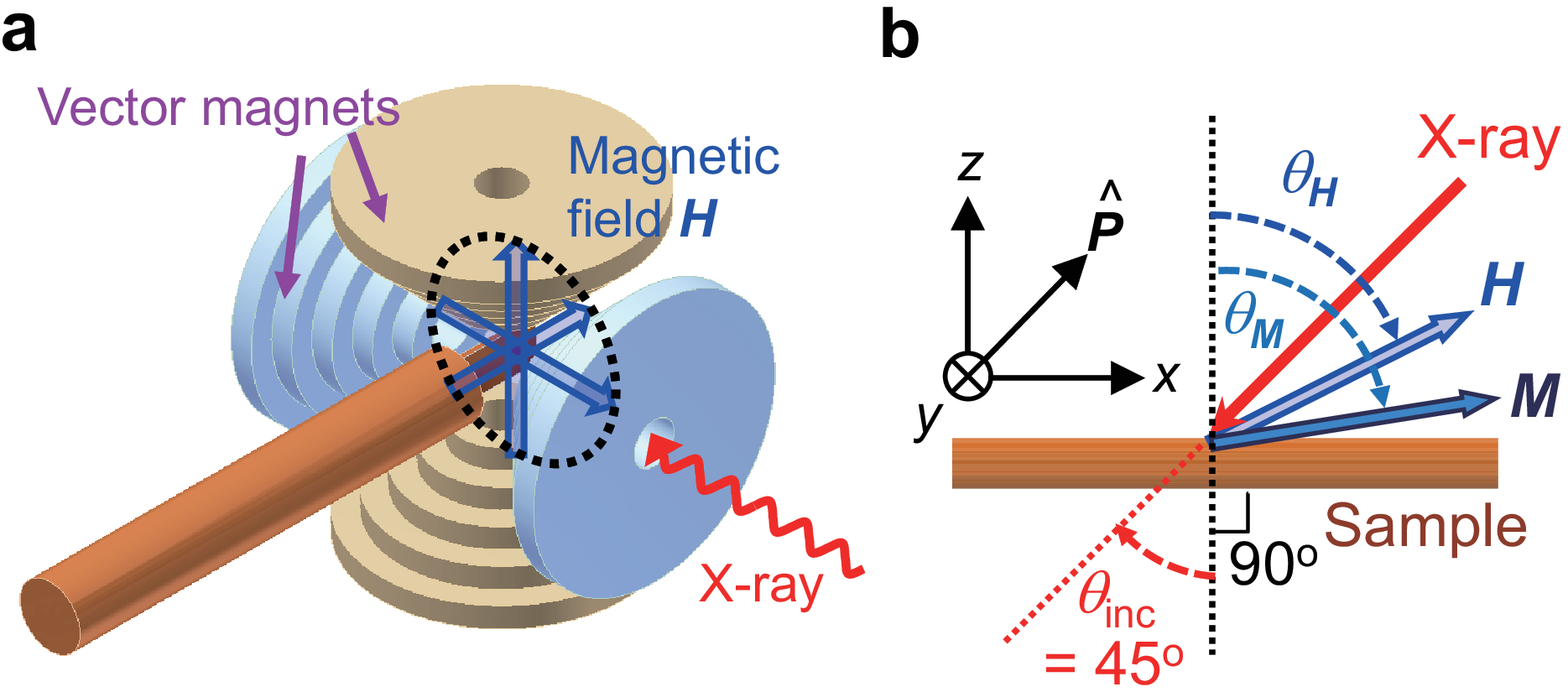} 
\caption{ }
\end{figure*}
\begin{figure*}
\includegraphics[width=16cm]{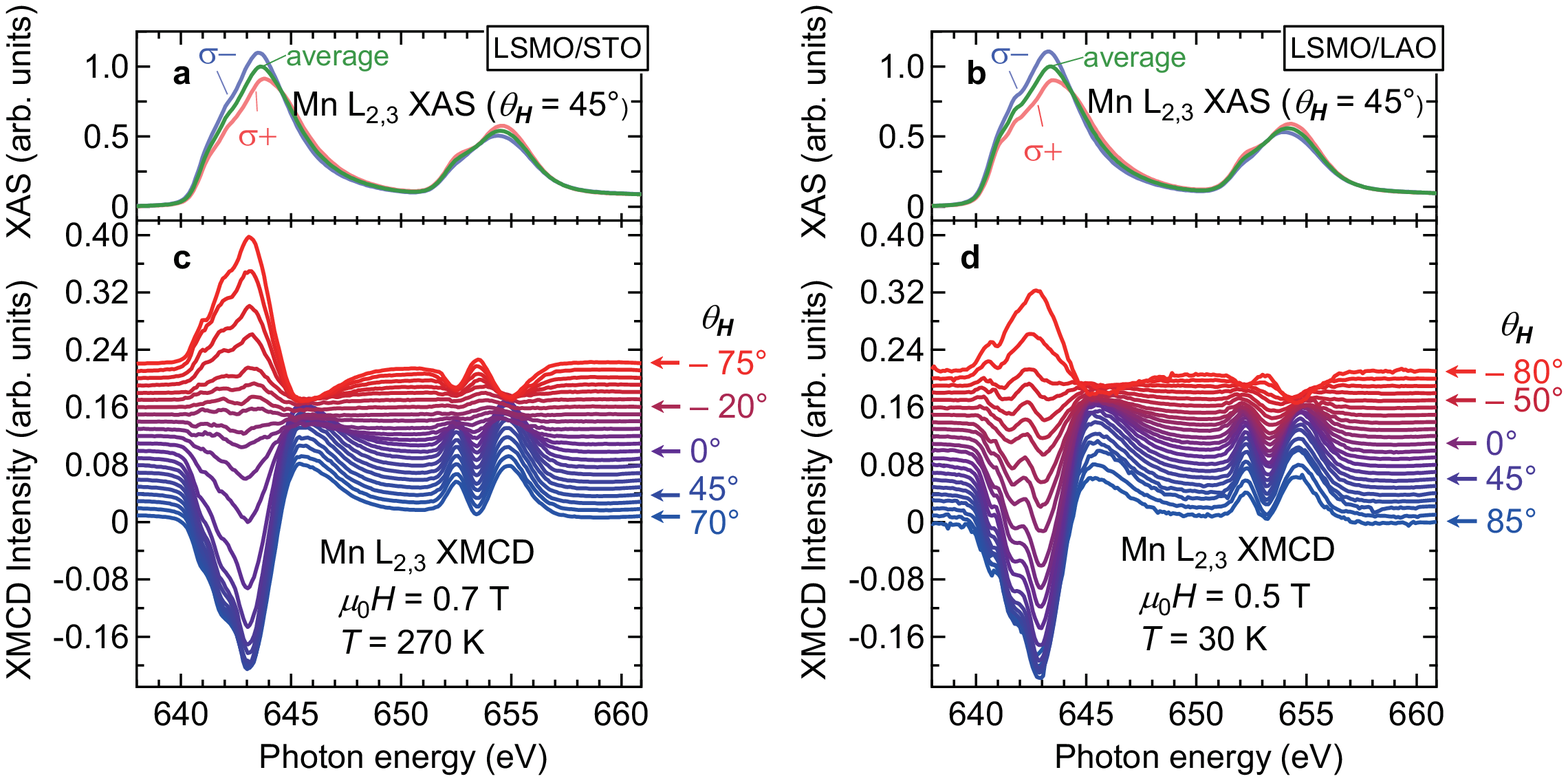} 
\caption{ }
\end{figure*}
\begin{figure*}
\includegraphics[width=16cm]{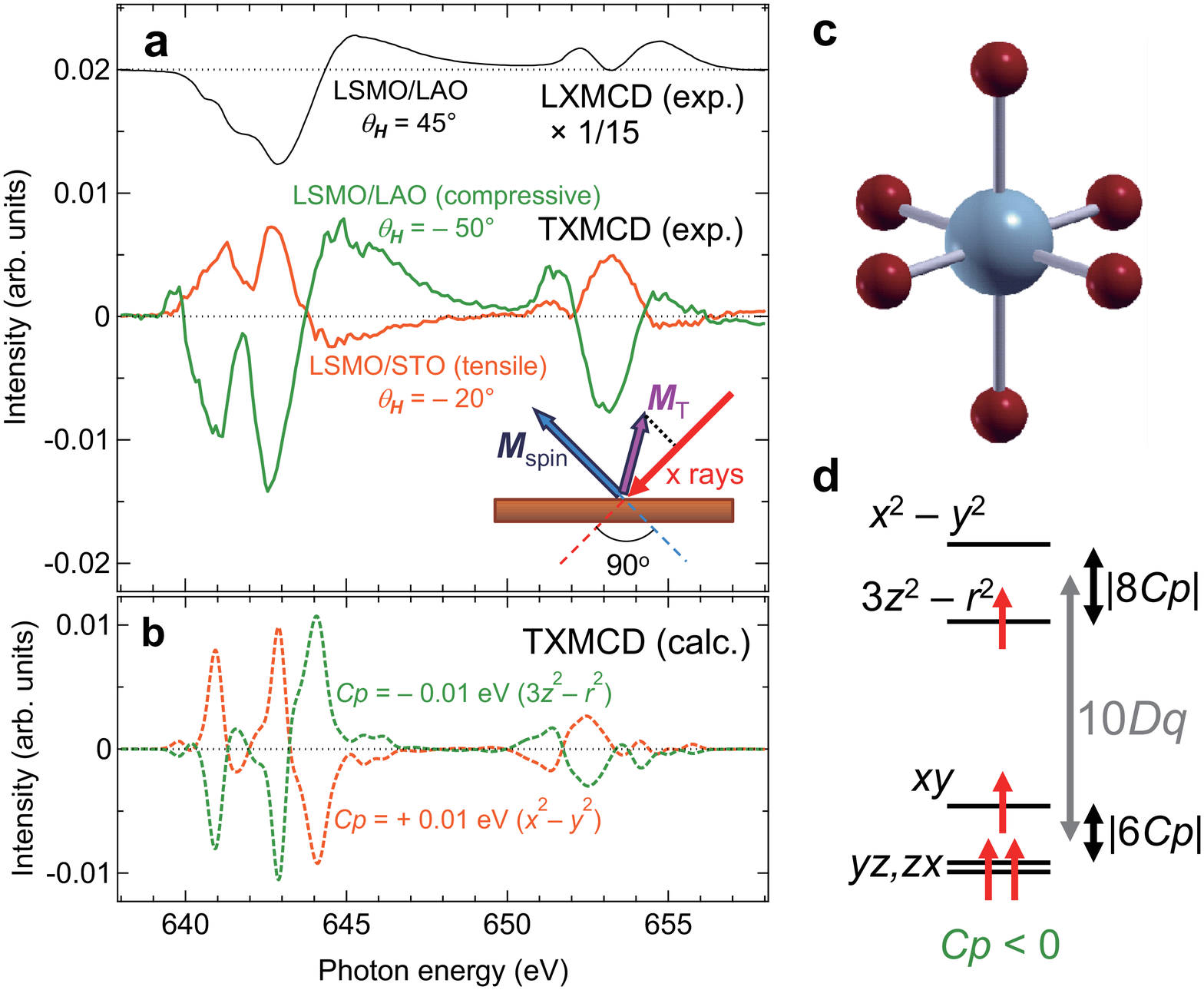}
\caption{ }
\end{figure*}
\begin{figure*}
\includegraphics[width=16cm]{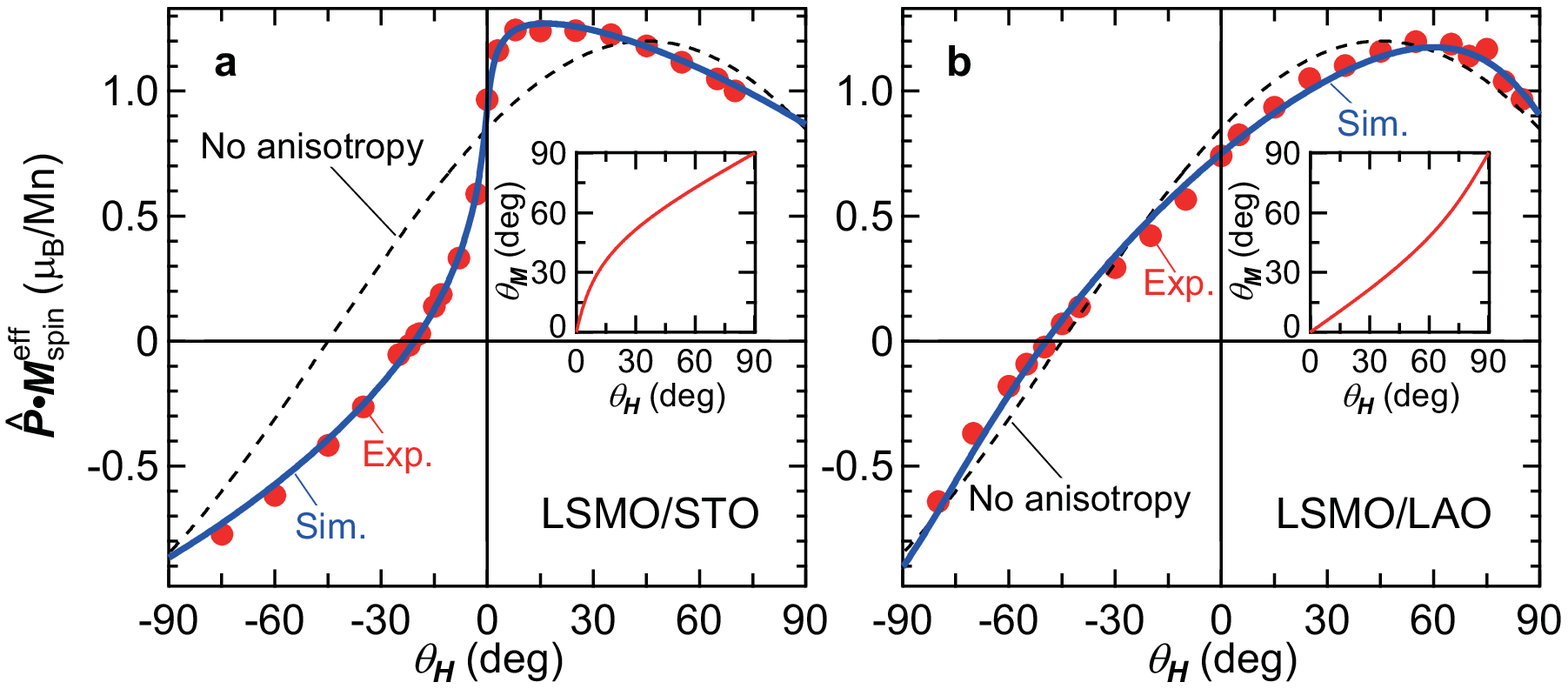}
\caption{ }
\end{figure*}

\clearpage

\begin{table*}[t]
\centering
\caption{
\textbf{Best-fit parameters for the simulated curves in Figs.\ \ref{adXMCD2}a and \ref{adXMCD2}b.} 
Errors have been estimated using the least squares method. 
Note that $K_u$ represents the MCA energy excluding shape magnetic anisotropy. 
}
\label{table1}
 \begin{ruledtabular}
 \begin{tabular}{cccc}
Substrate & $M_{\text{sat}}$ ($\mu_{\text{B}}$/Mn) & $K_u$ (kJ/m$^3$) & $(7/2)\langle Q_{zz} \rangle$   \\ \hline
  STO     & $1.255 \pm 0.007$                        & $-37.2 \pm 0.8$      & $+0.05 \pm 0.01$ \\ 
  LAO     & $1.206 \pm 0.014$                        & $+40.4 \pm 2.4$      & $-0.12 \pm 0.02$ \\ 
 \end{tabular}
 \end{ruledtabular}

\end{table*}

\end{flushleft}

\end{document}